%% file: 0main.tex
\documentclass[%
 reprint,
superscriptaddress,
showpacs,preprintnumbers,
nofootinbib,
 amsmath,amssymb, 
 aps,
 prd,
 longbibliography,
]{revtex4-1}

\usepackage{cancel}
\usepackage{accents}
\usepackage{mciteplus,slashed}
\usepackage{amssymb,cancel,amsmath}
\usepackage{dcolumn}
\usepackage{bm}
\usepackage[caption=false]{subfig}
\usepackage{appendix}
\usepackage{physics}
\usepackage{booktabs}
\usepackage{feynmp-auto}
\usepackage[T1]{fontenc}	
\usepackage{csvsimple}
\usepackage{hyperref}
\usepackage[section]{placeins}
\usepackage[capitalise]{cleveref}
\usepackage{booktabs}
\usepackage{graphicx}
\usepackage{mathrsfs}
\graphicspath{{Figures/}}

\AtBeginDocument{
\heavyrulewidth=.08em
\lightrulewidth=.05em
\cmidrulewidth=.03em
\belowrulesep=.65ex
\belowbottomsep=0pt
\aboverulesep=.4ex
\abovetopsep=0pt
\cmidrulesep=\doublerulesep
\cmidrulekern=.5em
\defaultaddspace=.5em
}
\usepackage[dvipsnames]{xcolor}
\usepackage[normalem]{ulem}
\usepackage{fontawesome} 

\input{Commands/ryansCommands.tex}

\input{Commands/flags.tex}
\begin{document}

\title{Luminous solar neutrinos I: Dipole portals}

\author{Ryan Plestid}
\email{rpl225@uky.edu}
\affiliation{Department of Physics and Astronomy, University of Kentucky  Lexington, KY 40506, USA}
\affiliation{Theoretical Physics Department, Fermilab, Batavia, IL 60510,USA}
\preprint{FERMILAB-PUB-20-521-T-V}

\begin{abstract}
{\centering{\href{https://github.com/ryanplestid/luminous-solar-nu}{\large\color{BlueViolet}\faGithub}}  \\}  
     Solar neutrinos upscattering inside the Earth can source unstable particles that subsequently decay inside large volume detectors (e.g.\ neutrino experiments). Contrary to naive expectations, when the decay length is much shorter than the radius of the \emph{Earth} (rather than the length of the detector), the event rate is independent of the decay length.  In this paper we study a neutrino dipole portal (transition dipole operator) and show that existing data from Borexino and Super Kamiokande probes previously untouched parameter space in the 0.5--20 MeV regime, complementing recent cosmological and supernova bounds.  We  discuss similarities and differences with luminous dark matter and comment on future prospects for analogous signals stemming from atmospheric neutrinos. A companion paper explores an analogous mass-mixing portal. 
\end{abstract}

 \maketitle 
 
 \section{Introduction \label{Introduction}}
 The Earth is constantly pelted with neutrinos from the Sun (among other sources), and dark matter from our local galaxy; both can serve as a resource with which to search for new physics. The canonical strategy, as it pertains to dark matter, is to detect particles directly via elastic or inelastic scattering within a detector, however this strategy applies equally well to new physics coupled via some neutrino-portal. For example, electron recoil data can be used to constrain both elastic  and transition neutrino dipole moments, both of which have attracted recent interest as potential explanations of the XENON1T excess \cite{Shoemaker:2020kji,Aprile:2020tmw,Babu:2020ivd,AristizabalSierra:2020zod,Miranda:2020kwy} (see also \cite{Giunti:2014ixa} for a review) . More broadly speaking, a transition dipole operator can serve as  ``dipole portal'' to a dark sector \cite{Aparici:2009fh,Magill:2018jla,Brdar:2020quo} and is described by the interaction Lagrangian\footnote{We use the variable $d$ rather than $\mu_\nu$ throughout this work. A quick conversion factor that is convenient for comparing results across conventions is $d \times 6.75~\text{MeV} = \mu_\nu/\mu_B$ with $\mu_B$ the Bohr magneton.} 
 \begin{equation}\label{dipole-portal}
    \mathcal{L}_\text{int}\supset d_{a} \bar{N}_R \sigma_{\mu\nu} \nu_L^{(a)} F^{\mu\nu} = \frac{\mu_\nu^{(a)}}{2}  \bar{N}_R \sigma_{\mu\nu} \nu_L F^{\mu\nu} ~,
\end{equation}
where $\nu_L$ is the Standard Model left-handed neutrino, and $F_{\mu\nu}$ is the electromagnetic field strength. The index $a=e,\mu,\tau$ denotes the neutrino flavour and $N_R$ is a (possibly Dirac or Majorana) sterile neutrino. The dipole operator, $d_a$,  can generically be flavour dependent however since the solar neutrino flux contains all three flavors the sensitivity derived here is only midly flavor dependent. This model was proposed as a phenomenological explanation of the MiniBooNE and LSND anomalies \cite{Gninenko:2010pr}, and subsequently studied in the context of radiative muon capture \cite{McKeen:2010rx}, supernovae cooling \cite{Magill:2018jla}, collider searches \cite{Magill:2018jla}, double bangs at ICECUBE \cite{Coloma:2017ppo},  fixed-target facilities \cite{Gninenko:1998nn,Masip:2012ke,Alvarez-Ruso:2017hdm,Magill:2018jla}, low-recoil dark matter detectors \cite{Shoemaker:2018vii} and cosmology of the early universe \cite{Brdar:2020quo}.  This paper is the first in a series of two that investigates the physics potential for detecting upscattered unstable particles sourced by solar neutrinos inside the Earth with a neutrino dipole portal serving as a useful benchmark example. In \cite{Plestid:2020ssy} we consider a mass-mixing portal whose upscattering and decay phenomenology is sufficiently different to warrant an isolated treatment.

 The parametric sensitivity to the dipole operator, $d$, of a search strategy that uses a flux of incident neutrinos \emph{directly} is straightforward. The rate of scattering events inside a detector is proportional to the flux times the cross section $\Phi \times \sigma$ multiplied by the number of targets inside the detector's fiducial volume. The flux of incident neutrinos is independent of $d$, while the detection cross section scales quadratically with the dipole operator, $\sigma \sim d^2$. The signal rate, $R$, therefore scales as $R\sim d^2$. 
 \begin{figure}
    \includegraphics[width=\linewidth]{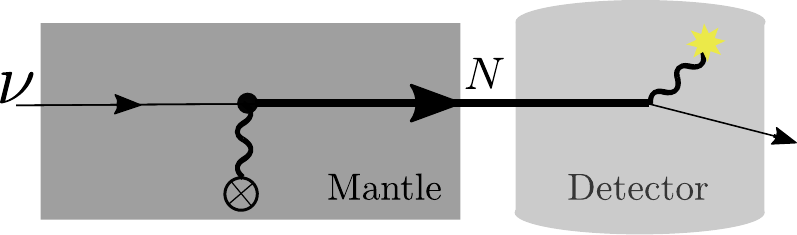}
    \caption{Upscattering of a solar neutrino inside the Earth's mantle (or core, or crust) leads to a ``heavy'' $O($MeV$)$ sterile neutrino, $N$. In the weak coupling limit $N$ is long-lived, and can propagate long distances eventually decaying inside a large-scale detector. For a neutrino-dipole portal this leads to a deposited photon with an energy of order $\sim$ 100 keV - 10 MeV.  \label{cartoon}}  
 \end{figure}
 Another possibility is to make use of astrophysical fluxes \emph{indirectly} by leveraging their ability to \emph{produce} a flux of new physics particles. For instance, cosmic rays interacting in the upper atmosphere can produce long-lived particles \cite{Arguelles:2019ziu} in much the same way that dedicated fixed target facilities (e.g.\ NA62 \cite{Iacobuzio:2712960}, SHiP \cite{SHiP:2018xqw} etc.) can. In both cases $pp$ and/or $pA$ collisions source mesons, which promptly decay, ultimately producing a flux of light, long-lived, new physics particles e.g.\  heavy neutral leptons \cite{Arguelles:2019ziu}, light dark matter \cite{Alvey:2019zaa},  or millicharged particles \cite{Plestid:2020kdm}. Such an indirect production mechanism costs an extra two powers of the coupling constant, such that for a neutrino dipole-portal the flux itself scales as $\Phi\sim d^2$ leading to an event rate that scales as $R\sim d^4$. These indirect fluxes are therefore expected to provide meaningful sensitivity to new physics only at moderately small couplings.

 A seemingly innocuous twist on this latter scenario is to consider an indirect flux of new particles that decay within a detector rather than scattering against its constituents. Consider a detector with a characteristic length scale $\Ldet$, and a particle with a decay length, $\Ldec$, that satisfied $\Ldec\gg \Ldet$. A long-lived particle is unlikely to decay inside a finite sized detector, since the probabilitty of decay within a distance $\Ldet$ is, $P_\text{dec}=1-\exp[-\Ldet/\Ldec]$ and for $\Ldec\gg \Ldet$ this scales as $P_\text{dec} \approx \Ldet/\Ldec\sim d^2$. Naively, therefore, the signal rate in this scenario scales as $R\sim d^4$. 
 
 The parametric dependence of the signal changes dramatically, however, if the indirect flux is sourced by upscattering inside the Earth and the decay length satisfies $\Ldec \ll R_\oplus$ with $R_\oplus$ the radius of the Earth. In this case, the upscattered flux that arrives at the detector is proportional to $\Ldec$ such that $\Phi \times P_\text{dec}$ is independent of $\Ldec$ at leading order in $\ell/\Ldec$. This can be understood as the effective column density of targets growing with $\Ldec$ in such a way as to precisely cancel the $1/\Ldec$ penalty arising from the rarity of decays within the detector. This effect persists until it is saturated by the boundaries of the Earth after which, rather than being suppressed by a factor that is $O(\Ldet/\Ldec)$ the rate is instead suppressed by a factor that is $O(R_\oplus/\Ldec$).
 
 We can make our discussion more concrete by considering a flux of incident particles on a thick slab of material of length $L_\text{slab}$ which terminates in a detector as depicted in \cref{cartoon}. If we consider an infinitesimally thin slice of the slab (thickness $\dd z$) then the flux of long-lived particles, $N$, arriving at the front detector is $\dd \Phi_N = \Phi_{\nu\odot} \overline{n}_A \sigma_{\nu\rightarrow N}\e^{-z/\Ldec} \dd z$, where $z$ is the distance from the slice to the detector, $\overline{n}_A$ is the number density of upscattering targets and $\sigma_{\nu\rightarrow N}$ is the upscattering cross section for $\nu A \rightarrow N A$. Integrating over $z$ we find the flux at the detector is given by 
 \begin{equation}\begin{split}
    \Phi_N&=  \Phi_{\nu\odot} \overline{n}_A \sigma_{\nu\rightarrow N}\int_0^{L_\text{slab}} \e^{-z/\Ldec} \dd z \\
    &= \Phi_{\nu\odot} [\overline{n}_A \Ldec] \sigma_{\nu\rightarrow N} (1-\e^{-L_\text{slab}/\Ldec})~,
\end{split} \end{equation}
 where the quantity in the square braces can be interpreted as the effective column density of scatterers along the line of sight. The rate of decays within the detector will be proportional to the product of this flux, the area of the detector, and the probability of decaying within it
 \begin{equation}\begin{split}
    R_\text{dec}& = \Phi_N A_\text{det} (1-\e^{-\Ldet/\Ldec})\\
    &\approx \Phi_{\nu\odot} V_\text{det} \overline{n}_A  \sigma_{\nu A\rightarrow N A}~,
 \end{split}\end{equation}
 where we have assumed $L_\text{slab} \gg \Ldec\gg \Ldet$. This can be compared to the rate of quasi-elastic scattering $\nu X\rightarrow X \nu$ signal events from the \emph{direct} flux of neutrinos 
 \begin{equation}\begin{split}
    R_\text{el}& = \Phi_{\nu\odot} V_\text{det} n_X  \sigma_{\nu X \rightarrow N X}~.
 \end{split}\end{equation}
We have included the label $X$, because for scattering events to be visible inside the detector, their energy deposition must be observable as $X$-recoil energy. Low-energy nuclear recoils are difficult to observe as compared to electron recoils, which means that $\nu e\rightarrow N e$ scattering often provides better sensitivity. Upscattering off of electrons, however, has a much smaller cross section than $\nu A \rightarrow N A$. In upscatter-decay scenarios, the nuclear recoil of a target inside the Earth does not need to be detected, and so $\nu A\rightarrow N A$ scattering is an ever-present production mechanism which dominates over $\nu e \rightarrow N e$.

\begin{figure}
    \includegraphics[width=\linewidth]{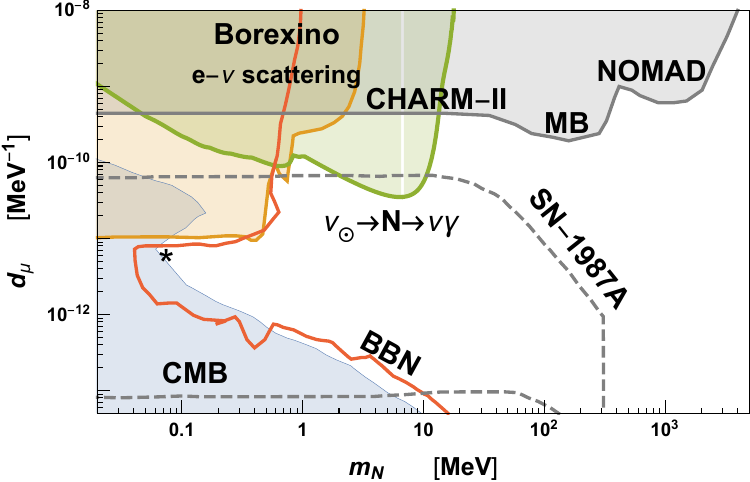}
    \caption{Constraints on a muon-only dipole coupling $d_\mu$, from cosmology (BBN and CMB) \cite{Brdar:2020quo},  SN-1987A \cite{Magill:2018jla}, Borexino $e\nu$ scattering data \cite{Brdar:2020quo,Agostini:2018uly,Borexino_data}, CHARM-II $e\nu$ scattering data \cite{Coloma:2017ppo,Geiregat:1991md}, MiniBooNE \cite{Magill:2018jla,AguilarArevalo:2007it}, and NOMAD \cite{Gninenko:1998nn,Altegoer:1997gv}; the last viable parameter space \cite{Brdar:2020quo} to explain the XENON1T excess is indicated with a star. This work $(\nu_\odot \rightarrow N\rightarrow \nu\gamma)$ is shown in green with a solid (dashed) line for $\alpha=-1$ (($\alpha=0$) corresponding to a maximally CP-violating Dirac  (Majorana) $N$. Constraints were obtained by multiplying the solar neutrino flux by $P_{e\mu}(E_\nu)$ \cite{Akhmedov:2004rq,Harnik:2012ni,Vedran-PC}.  \label{muon-only}} 
 \end{figure}

 Therefore, despite the upscattered flux being indirect (i.e.\ sourced by scattering) the event rate is parametrically identical to direct detection (scaling as $R\sim d^2$ rather than $R\sim d^4$). Furthermore, one can clearly see a number of avenues via which event rates from upscattered long-lived particles can supersede those of direct elastic recoil:
 \begin{enumerate}
    \item While both event rates are proportional to the volume of the detector, $R_\text{dec}$ scales with the density of upscattering targets \emph{inside the Earth}. The interior of the Earth tends to be 3-12 times more dense than detector material.
    \item The cross sections entering the two expressions are different, and the upscattering cross section may be much larger. For example  $\nu e \rightarrow N e$ (detection) has a much smaller cross section than $\nu A \rightarrow N A$ (upscattering). 
    \item Direct detection may be kinematically disfavoured as emphasized in the literature surrounding luminous dark matter \cite{Feldstein:2010su,Pospelov:2013nea,Eby:2019mgs}.  
 \end{enumerate}
 Even when $\nu X\rightarrow N X$ scattering is kinematically allowed, points 1.\ and 2.\ can make the decay event rate much larger than the elastic scattering event rate. 
 
 The rest of this paper is dedicated to investigating the sensitivity of large-scale detectors to a dipole portal for $m_N \leq 18.8$ MeV, the cut-off of the solar-neutrino flux. Our results are summarized and shown in context with other constraints for a muon-only dipole portal in \cref{muon-only}; constraints derived in this paper for $d_e$ and $d_\tau$ are broadly similar, while constraints based on fixed-target facilities either weaken or disappear entirely for $d_e$ and $d_\tau$ because the flux considered is predominantly composed of muon neutrinos. 
 Solar neutrinos offer a robust probe of low-mass sterile neutrinos independent of neutrino flavor. 
 
 The rest of the article is organized as follows: In \cref{Upscattering} we discuss the upscattering of solar neutrinos into unstable right-handed neutrinos via the dipole portal. Next, in \cref{Decay} we discuss decay properties including the difference between Majorana and Dirac decays \cite{Shrock:1982sc,Pal:1981rm,Balantekin:2018ukw}. Then, in \cref{Signals}, we discuss how these two pieces of phenomenology conspire to produce photon spectra at large-scale detectors on Earth, the details of which are determined by the geometry of the Earth relative to the Sun, and the latitude of the detector.  In \cref{Constraints} we use the photon spectra derived in \cref{Signals} to set limits on a dipole portal couplings using Borexino \cite{Agostini:2018uly,Borexino_data} and Super-Kamiokande data \cite{Abe:2016nxk}. We conclude with \cref{Conclusions} where we summarize our main results, emphasize the overlap between luminous dark matter searches and luminous solar neutrino searches, and comment on qualitative effects that are important for future dedicated analysis. 

\begin{figure}
    \includegraphics[width=0.9\linewidth]{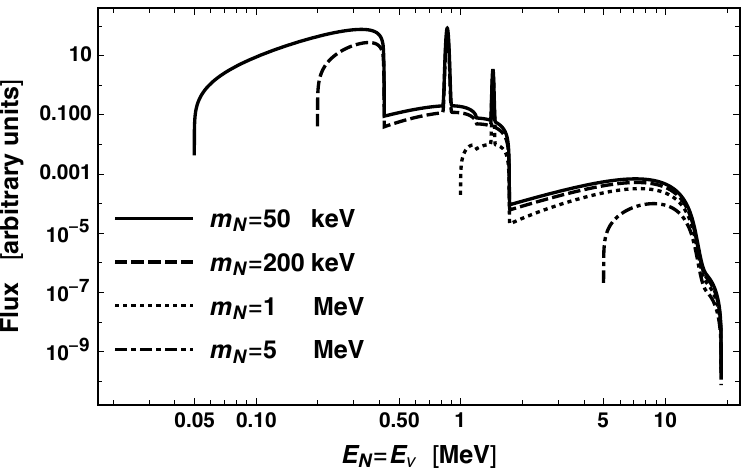}
    \caption{Shape of the sterile neutrino flux,  $\Phi_N \propto  \Phi_{\nu\odot} \times \sigma_{\nu\rightarrow N}^\leadsto$,  for various choices of $m_N$ \emph{c.f.} \cref{Solar-Flux}. For a flavor dependent dipole portal $d_a$ the flux should be multiplied by the survival/transition probability,  $P_{ea}(E_\nu)$. \label{N-spec}}
\end{figure}
\section{Dipole-portal upscattering \label{Upscattering}}

 Solar neutrinos never exceed $20$ MeV in energy, and this means that their (hypothetical) electromagnetic interactions with nuclei fall firmly in the coherent regime $F(Q^2)\approx 1$. Moreover, the four-momentum transfer is limited by $E_\nu$, and so the recoil energy of a nucleus, $A$, never exceeds $T_A^\text{max} \sim (20 \text{MeV})^2/m_A \lesssim 10$ keV. Therefore, neutrino-up-scattering on nuclei can be reliably approximated by considering the nucleus as a fixed and static charge density sourcing a Coulomb field. Within this approximation $E_N=E_\nu$ by energy conservation  and the matrix element for up-scattering is given by
\begin{equation}
    \left\langle|\mathcal{M}|^2\right\rangle= \frac{4 d^2 (Z e)^2 }{t}\qty[ 4 E_\nu^2 - m_N^2 +m_N^4/t] ~,
\end{equation}
where flavour indices have been suppressed for brevity's sake. Using $\dd \sigma =\frac{1}{16\pi^2}\left\langle|\mathcal{M}|^2\right\rangle \dd \Omega$ (appropriate for potential scattering), noting that $\dd t = 2 E_\nu^2\dd \cos\theta$, and integrating over the azimuthal angle we have 
\begin{equation}\begin{split}
    \dv{\sigma}{t}&=\frac{4 d^2 Z^2(4\pi \alpha) }{16\pi} \frac{1}{t}\qty[ 4 E_\nu^2 - m_N^2 +m_N^4/t]\\
    &=\frac{(d Z\alpha)^2}{t} \qty[ 4 E_\nu^2 - m_N^2 +m_N^4/t]~.
\end{split}\end{equation}
This is logarithmically enhanced to prefer forward scattering such that it is a good approximation to treat the resultant ``beam'' of steriles produced via solar neutrino upscattering to be parallel with the solar neutrino flux. We restrict upscattering to ``forward'' angles satisfying $\cos\theta > 0$ such that (for $Q^2=-t$) 
\begin{align}
    Q^2_\text{min} &= (P_\nu - P_N)^2 
    = \qty(E_\nu- \sqrt{E_\nu^2 -m_N^2})^2 \\
    Q^2_\text{max} &= P_\nu^2 + P_N^2= 2 E_\nu^2 -m_N^2~. 
\end{align}
and clearly in the $E_\nu\gg m_N$ limit this give the result $Q_\text{min}= \tfrac12 m_N^2/E_\nu$ and $Q_\text{max}\approx 2E_\nu$. We can then calculate the rate of ``forward'' scattering (denoted by the superscript $\leadsto$) as
\begin{widetext}
\begin{equation}\begin{split}\label{cross-section}
  \sigma_{\nu\rightarrow N}^\leadsto &= 4 Z^2\alpha d^2\qty[
    \log\qty(\frac{Q^2_\text{max}}{Q^2_\text{min}}) + \frac14 \frac{m_N^4}{E_\nu^2 Q_\text{min}^2} - \frac14 \frac{m_N^2}{E_\nu^2}\log\qty(\frac{Q^2_\text{max}}{Q^2_\text{min}}) -\frac14 \frac{m_N^4}{E_\nu^2 Q_\text{max}^2} ]\Theta(E_\nu-m_N)\\
  &\approx 4Z^2\alpha d^2 \qty[\log\qty[\frac{(2E_\nu)^4}{m_N^4}]  + 1-\log2   +  O(\epsilon^2\log\epsilon) ] \Theta(E_\nu-m_N)~.
\end{split}\end{equation}
\end{widetext}
The terms in the top line have been organized based on their scaling with respect to $\epsilon = m_N/E_\nu$. At logarithmic accuracy in $\epsilon$, we therefore have $\sigma \sim 16 Z^2 \alpha d^2 \log(2E_\nu/m_N)$ with corrections being of order $\sim \text{few}~\%$ for $\epsilon<1/2$. The $Z^2$ enhancement means that scattering from nuclei will always dominate over scattering from electrons. Keeping only the leading-logarithm and taking $d=1.97\cdot10^{-9}$ MeV$^{-1}$ [a convenient choice for decay-length purposes \emph{c.f.} \cref{dec-length}]
\begin{equation}
    \sigma_{\nu\rightarrow N}^\leadsto\approx [1.76
    \cdot 10^{-40}~\text{cm}^2]~ Z^2 \log(2E_\nu/m_N)~.
\end{equation}
For $E_\nu \gtrsim  m_N$ (near threshold) one must use the full expression in the first line of \cref{cross-section} and the resultant flux is shown for a few benchmark masses in \cref{N-spec}.

The flux of steriles $\Phi_N$ emerging from an infinitesimal slab of thickness $\dd z$ with target density $\overline{n}_A$ is then given by 
\begin{equation}
    \dd \Phi_N(E_N) =    
    (\overline{n}_A \times \dd z) \sigma_{\nu\rightarrow N}^\leadsto \Phi_{\nu\odot} (E_\nu) ~.
\end{equation}
This flux, unsculpted by the energy dependent decay length, is plotted in \cref{N-spec}.

For flavor dependent couplings oscillation effects should be included in the solar neutrino spectrum and the result should be summed over flavors appropriately weighted by $d_a^2$. This leads to, at most, an $O(1)$ modification of the constraints and we neglect it in the discussion hereafter, but include it in our flavor-dependent results \cref{muon-only}.

\begin{figure}
    \includegraphics[width=0.9\linewidth]{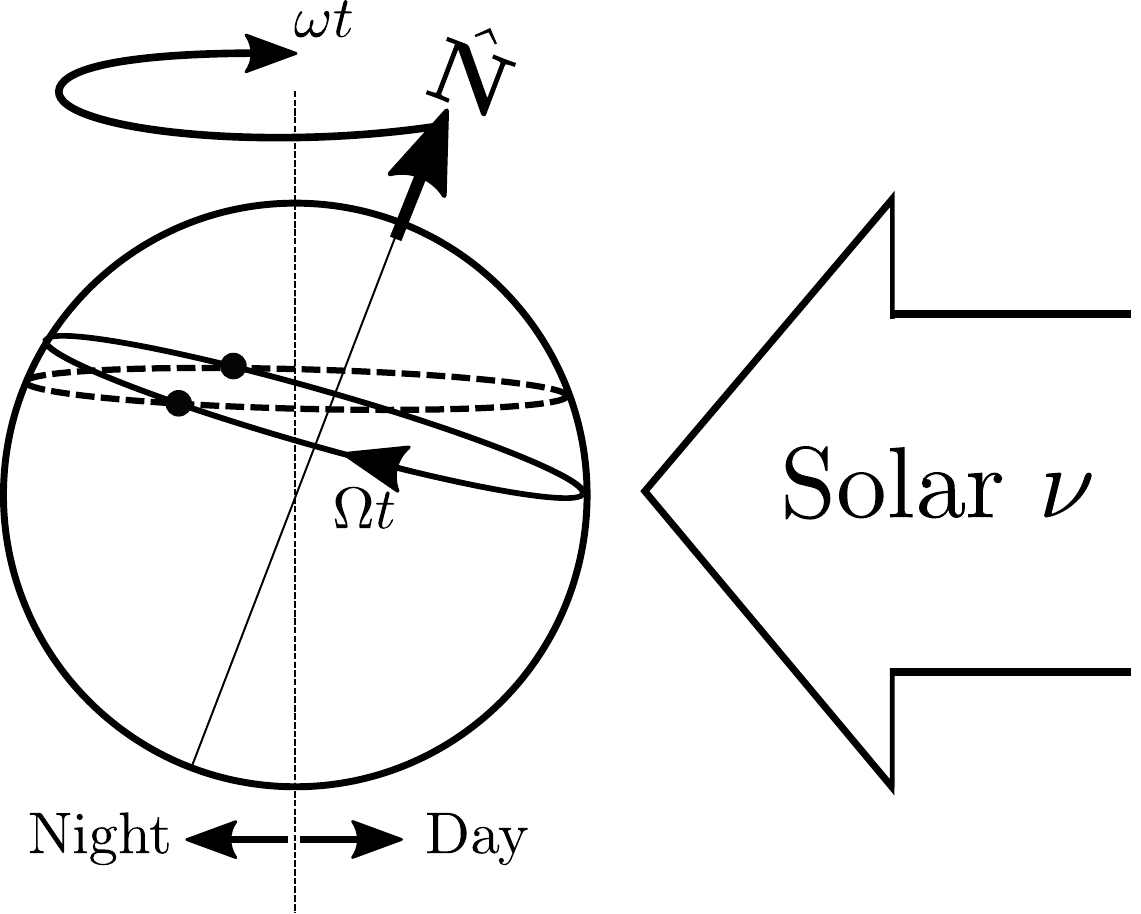}
    \caption{Time dependent motion of a detector relative to the solar neutrino flux. A detector at a fixed latitude traces out a path depicted by the solid line [$\Omega -\omega = 2\pi/(1~\text{day}) $ and $\omega= 2\pi/(365~\text{days})$].  \label{Earth-sketch}}
\end{figure}

\section{Radiative sterile decay \label{Decay} }

Once the flux of $N$'s has been determined, the resultant spectral shape of the photons from $N\rightarrow \nu \gamma$ can be calculated. The shape is somewhat model dependent, being determined by the Majorana vs Dirac nature of $N$ and the level of CP-violation which is present \cite{Balantekin:2018azf,Balantekin:2018ukw},  which determines the angular dependence of the differential decay rate in the rest frame of $N$
\begin{equation}
    \dv{\Gamma}{\cos\theta}\propto (1+ \alpha \cos\theta)\quad \alpha\in [-1,1]~.
\end{equation}
\vspace{6pt}

A Majorana $N$ has $\alpha=0$, whereas a Dirac $N$ (naively favoured to suppress dipole contributions to neutrino textures \cite{Magill:2018jla}) can have $\alpha\neq0$ for a combination of magnetic and electric dipole moments \cite{Shrock:1982sc,Pal:1981rm,Balantekin:2018azf,Balantekin:2018ukw}. For illustration we include $\alpha=0$ and $\alpha=-1$.  In the lab frame, for an $N$ with energy $E_N$, this leads to 
\begin{equation}
    \dd \Gamma / \dd E_\gamma =
    \begin{cases}
    \text{box}(E_\gamma,E_N) & \alpha=0\\
    \text{tri}(E_\gamma,E_N) & \alpha=-1
    \end{cases}~,
\end{equation}
where 
\begin{align}
    \text{box}(E_\gamma)&=\frac{\Theta(E_\gamma-E_\gamma^{(-)})   \Theta(E_\gamma^{(+)}-E_\gamma)}{E_\gamma^{(+)}-E_\gamma^{(-)}} \\
    \text{tri}(E_\gamma)&=2\frac{E_\gamma^{(+)}-E_\gamma}{E_\gamma^{(+)}-E_\gamma^{(-)}} \text{box}(E_\gamma)\\
    E_\gamma^{(\pm)}&= \frac{E_\nu}{2} \qty( 1 \pm \sqrt{1- \frac{m_N^2}{E_\nu^2}})~.
\end{align}
The photon energy spectrum is then obtained by integrating over $E_N$ weighted by the flux of $N$. For example for the $\alpha=-1$ case we have 
\begin{equation}\begin{split}
    \dv{R(E_\gamma)}{z} = (1-&\e^{-\Ldet/\Ldec})(1-\e^{-z/\Ldec})\times \\
    & \int \dd E_N \dv{\Phi_N(E_N)}{z} \text{tri}(E_\gamma,E_N)~.
\end{split}\end{equation}
Taking the density to be constant, $\overline{n}_A(z)=\overline{n}_A$, integrating over $z$, treating $\Ldet\ll \Ldec$, and multiplying by the size of the detector we arrive at 
\begin{widetext}
\begin{equation}\label{Slab-Photon-Rate}
    R(E_\gamma) = V_\text{det} \overline{n}_A  
    \int_{m_N}^{18.8~\text{MeV}} \dd E_N (1-\e^{-L_\text{slab}/\Ldec}) \Phi_{\nu\odot} (E_\nu = E_N)  \sigma_{\nu\rightarrow N}^\leadsto (E_N)  \times \begin{cases}
    \text{box}(E_\gamma,E_N) \\
    \text{tri}(E_\gamma,E_N)
    \end{cases}~.
\end{equation}
\end{widetext}
This form is applicable to a spatially varying density profile if we make the replacement $\overline{n}_A (1-\e^{-L_\text{slab}/\Ldec}) \rightarrow \frac{1}{\Ldec}\int \dd z  \overline{n}_A(z) (1-\e^{-z/\Ldec})$. 

We have implicitly assumed that $\Phi_{\nu\odot}(E_\nu)$ is independent of $z$. This assumption can be violated if $d$ is flavour non-universal ($d_e\neq d_\mu \neq d_\tau$) whereupon neutrino oscillations spoil this picture in principle. For solar neutrinos, however, the incident flux is a statistical mixture of  $O(1)$:$O(1)$:$O(1)$ for $\nu_e$:$\nu_\mu$:$\nu_\tau$. Constraints from solar neutrino fluxes are therefore insensitive to the flavour dependence of the coupling constants aside from $O(1)$ effects due to e.g. only $1/3$ of the solar flux contributing to upscattering.

Omnipresent in our current discussion is the decay length $\Ldec$ whose relative size determines whether our signal scales as $R\sim d^2$ or $R\sim d^4$. The lifetime of $N$ is given by 
\begin{equation}
    \Gamma = \sum_a \frac{d_a^2 m_N^3}{4\pi} \times\begin{cases} 1 & \text{Dirac} \\
    2 & \text{Majorana}\end{cases}, 
\end{equation}
and its decay length, $\Ldec= \gamma \beta \tau$ is therefore 
\begin{equation}\label{dec-length}
\begin{split}
    \Ldec& = \frac{4\pi}{\sum_a d_a^2 m_N^3} \frac{E_N}{m_N}\sqrt{1-\frac{m_N^2}{E_N^2}}\\
    &= R_\oplus \qty[\frac{1.97 \times 10^{-9} ~\text{MeV}^{-1} }{d_\text{eff}}]^2  \qty[\frac{1~\text{MeV}}{m_N}]^4\\
    &\hspace{45pt}\times \qty[\frac{E_N}{10~\text{MeV}}] \sqrt{\frac{1-m_N^2/E_N^2}{0.99}}~. 
\end{split} 
\end{equation}
where $d_\text{eff}=\sum_a d_a^2$.

It turns out that the values of $d$ and $m_N$ that lead to a roughly Earth-scale decay length are right at the boundary of currently unexplored parameter space.  
\begin{figure}
    \includegraphics[width=\linewidth]{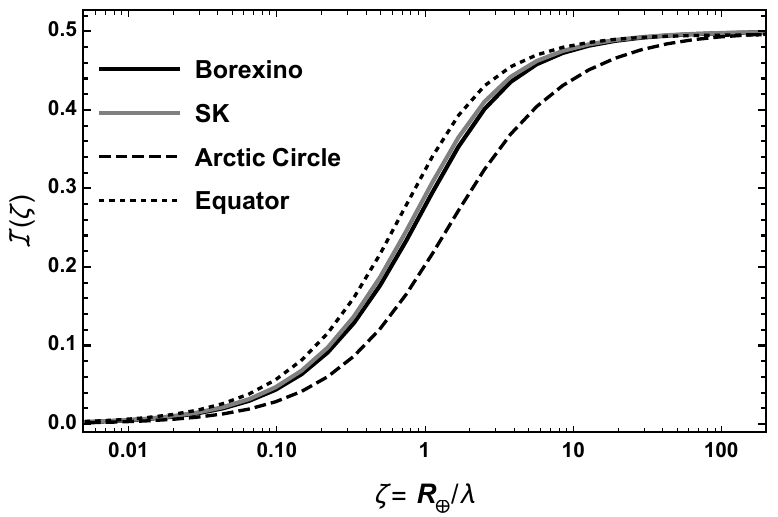}
    \caption{$\mathcal{I}(\zeta)$ (averaged over a year) for different latitudes.  Generally the closer a detector is to the equator the more advantageous its sensitivity will be, however this effect is relatively small provided we restrict possible detector sites to lie between the Arctic Circle and the Antarctic Circle. \label{I-fig}  }
\end{figure}

\section{Signals at large scale detectors \label{Signals}}

Unlike the simple picture presented in the introduction, the ``slab'' of Earth that is traversed by solar neutrinos in transit to a terrestrial detector is time dependent. This is obvious in that during the day the solar neutrinos pass through the Earth's crust whereas at night the majority of the line-of-sight density they encounter is the Earth's mantle. This can be visualized by working in a coordinate system where the solar flux is incident from the $\hat{x}$ direction. In this frame, the Earth rotates about its axis daily, and precesses about the $\hat{z}$ axis yearly as depicted in \cref{Earth-sketch}. 

Let us highlight a number of qualitative effects 
\begin{itemize}
    \item There is a stark day-night asymmetry with almost no signal during the day and all of the signal coming at night. The seasonal variation is less extreme being an $O(1)$ effect.   
    \item The neutrinos only pass through the core of the Earth\footnote{This is important for $\Ldec \gtrsim R_\oplus$. If $\Ldec \ll R_\oplus$ then particles upscattered in the core will decay before reaching the detector. }, which is composed of high-density and high-$Z$ material, during winter for detectors in the northern hemisphere.  Since $\Phi_N \propto Z^2 \overline{n}_A$, and the core is a high-density, high-$Z$ material this also introduces an important seasonal modulation. 
    \item The direction of the photons will be highly correlated with the zenith angle for highly boosted $N$'s (most of the spectrum).
\end{itemize}

Many of these issues are discussed in detail in \cite{Eby:2019mgs} for inelastic dark matter, which also has mostly-forward upscattering. For our present purposes we will focus on the time-averaged rate of photon deposition in a given experiment, and we therefore ignore details of seasonal and daily signal modulation. Furthermore, we consider a simplified model of the Earth that is designed to give a conservative estimate of the photon yield, while remaining analytically tractable.

We take the Earth to be a uniform sphere composed of one atomic species with $Z=Z_\text{eff}$ and number density $\overline{n}_A$, designed to mimic the average density of the Earth's mantle as shown in \cref{Earth-Material}. We assume that $\Ldec \gg $ few km, such that the detector is effectively on the surface. Each day the detector traces out the path labeled by the frequency $\Omega$ in \cref{Earth-sketch}. The length of the slab through which the neutrinos pass (shown as a dashed circle in \cref{Earth-sketch}) changes in time such that the factor of $(1-\e^{-L_\text{slab}/\Ldec})$ in \cref{Slab-Photon-Rate} is replaced by
\begin{equation}
    \mathcal{I}_i(\zeta) =\left\langle 1-\e^{-L(t)/\Ldec}\right\rangle_{i}~. 
\end{equation}
where $\langle ~\cdot ~ \rangle_i$ labels an average over the $i^\text{th}$ day of the year and we have introduced the dimensionless variable $\zeta=R_\oplus/\lambda$. 
The function $\mathcal{I}_i(\zeta)$ can be evaluated numerically, but also has simple analytic limits. For $\Ldec\ll R_\oplus$ we have the intuitive result that  
\begin{equation}
    \mathcal{I}_i(\zeta)\approx \frac{t^i_\text{night}}{24~\text{h}}~~~~\Ldec. \ll R_\oplus~,
\end{equation}
such that the rate of photons is independent of the decay length. Averaging over a full year's exposure would then yield a factor of $1/2$ because of the absence of signal during the day. In the opposite limit, where $\Ldec \gg R_\oplus$, the function is simply related to the average column density seen by the neutrinos 
\begin{equation}
    \mathcal{I}_i(\zeta)\approx \frac{1}{\Ldec}  \langle L(t) \rangle_{i}~~~\Ldec \gg R_\oplus~. 
\end{equation}

Making use of the seasonal modulation in signal could serve as a powerful tool for characterizing backgrounds and substantially improve the sensitivity of large terrestrial detectors as discussed in \cite{Eby:2019mgs} for the case of inelastic dark matter. 

For simplicity, we ignore seasonal variations and focus here on the total integrated rate, or equivalently the time averaged rate taken over a full year of run time. This has the advantage of being directly comparable to publicly released data from Borexino and SK. For this purpose we can define the average rate for the full live-time of an experiment as
\begin{align}
   \overline{R}(E_\gamma) =  \frac{1}{365}\sum_{i=1}^{365} R_i(E_\gamma)  ~,
\end{align}
where $R_i(E_\gamma)$ is defined the same way as in \cref{Slab-Photon-Rate} but with the replacement $(1-\e^{-L_\text{slab}/\Ldec})\rightarrow \mathcal{I}_i(\zeta)$. This leads immediately to the function 
\begin{equation}
    \mathcal{I}(\zeta) = \frac{1}{365}\sum_{i=1}^{365} \mathcal{I}_i(\zeta) ~,
\end{equation}
where we have introduced $\zeta(E_N,m_N,d) = R_\oplus/\Ldec $. The quantity $\mathcal{I}(\zeta)$ has a mild lattitude dependence as illustrated in \cref{I-fig}. With this function defined we can express the year-averaged differential rate of photon deposition, $\overline{R}(E_\gamma)$ (per unit time per unit energy) as 

\begin{widetext}
\begin{equation}
\label{Year-Avg-Photon-Rate}
    \overline{R}(E_\gamma) =  V_\text{det} \overline{n}_A 
    \int_{m_N}^{18.8~\text{MeV}} \dd E_N~ \mathcal{I}(\zeta) ~\Phi_{\nu\odot} (E_\nu = E_N)  ~\sigma_{\nu\rightarrow N}^\leadsto (E_N)  ~\times \begin{cases}
    \text{box}(E_\gamma,E_N) \\
    \text{tri}(E_\gamma,E_N) 
    \end{cases}~.
\end{equation}
In the limit where $\Ldec \ll R_\oplus$, but $\Ldec \gg h$ (with $h$ the overburden) we can replace $\mathcal{I}(\zeta) \rightarrow \tfrac12$ (half of the exposure is day-time where there is no signal\footnote{If $\Ldec \lesssim h$ then production in the Earth's crust (rather than the mantle) and there will be signal both day and night. }). If instead $\Ldec \gg R_\oplus$ (i.e.\ $\zeta \rightarrow 0$) then we can replace $\mathcal{I}(\zeta) \rightarrow \left\langle L(t)\right\rangle/\Ldec$ such that for the two limits we have 
\begin{align}
    &\dv{\overline{R}}{E_\gamma}= (104~\text{Hz}) \qty[\frac{V_\text{det} \overline{n}_A}{10^{30}}] \qty[\frac{Z_\text{eff}}{12}]^2 \cdot \begin{cases}
     \qty[\frac{d_\text{eff}}{1.97\cdot10^{-9}~\text{MeV}^{-1}}]^2 \cdot \frac{1}{2}\dv{R_\ll}{E_\gamma}   & \Ldec \ll R_\oplus\\
     \qty[\frac{d_\text{eff}}{1.97\cdot10^{-9}~\text{MeV}^{-1}}]^4 \cdot \qty[\frac{m_N}{1~\text{MeV}}]^4 \cdot \qty[\frac{\langle L(t) \rangle }{R_\oplus}] \dv{R_\gg}{E_\gamma} & \Ldec \gg R_\oplus
    \end{cases} \label{R-lims}\\
    &\dv{R_\ll}{E_\gamma}=\int_{m_N}^{18.8~\text{MeV}} \dd E_N ~\qty[\frac{\Phi_{\nu\odot}\overline{\sigma} }{10^{-30}~\text{Hz}}]  ~\times \begin{cases}
    \text{box}(E_\gamma,E_N) & \\
    \text{tri}(E_\gamma,E_N) &
    \end{cases}\label{Rll}\\
     &\dv{R_\gg}{E_\gamma}=\int_{m_N}^{18.8~\text{MeV}} \dd E_N \qty[\frac{10~\text{MeV}}{E_N}]\sqrt{\frac{0.99}{1-m_N^2/E_N^2}}
     \qty[\frac{\Phi_{\nu\odot}  \overline{\sigma}}{ 10^{-30}~\text{Hz}}]  ~\times \begin{cases}
    \text{box}(E_\gamma,E_N) & \\
    \text{tri}(E_\gamma,E_N) & 
    \end{cases}~,\label{Rgg}
\end{align}
\end{widetext}
where $\Phi_{\nu\odot}=\Phi_{\nu\odot}(E_\nu=E_N)$, and we have introduced a reference cross section,
\begin{equation}
\begin{split}
    \overline{\sigma}&= \sigma_{\nu \rightarrow N}^\leadsto \qty[\frac{1.97\cdot10^{-9} ~\text{MeV}^{-1}}{d}]^2 \qty[\frac{1}{Z^2}]\\
    &\sim O(1)~\times 10^{-40} \text{cm}^2~,
\end{split}
\end{equation}
that is independent of $d$ and $Z$, but depends on $m_N$ and $E_\nu$ as described in \cref{cross-section}.   

The function $\dd R_\ll/\dd E_\gamma$ is proportional to the differential rate of photons produced per target nucleus $A$, and is left unsculpted by the dependence of $\Ldec$ on $E_N$. In contrast, the function $\dd R_\gg /\dd E_\gamma$ \emph{is} sculpted by $\Ldec(E_N)\propto 1/P_N$, because the rate depends on the ratio $\langle L(t)\rangle/\Ldec$; up-scattered $N$'s with more energy are less likely to decay inside the detector. 

The most important feature of \cref{R-lims} is the transition between $R\sim d^2$ and $R\sim d^4$ scaling. The interpolation between these two limits is given by \cref{Year-Avg-Photon-Rate}, which can be safely used over the entire $m_N-d$ plane of parameter space. It is, however, instructive to consider experimental sensitivity in the two limits shown in \cref{R-lims}

Future dedicated analyses at large scale detectors should include core of the Earth, the radial dependence of the mantle density, photon spectral information, and make use of time-modulation to discriminate backgrounds from signal. At this level of detail it becomes mandatory to use \cref{Year-Avg-Photon-Rate} rather than \cref{R-lims} (especially in the cross-over region between the two limits). All results in this paper make use of \cref{Year-Avg-Photon-Rate}.

\section{New constraints on dipole portals \label{Constraints}}
\begin{figure}
    \includegraphics[width=\linewidth]{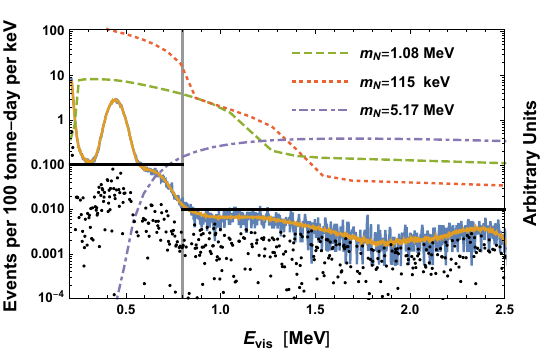}
    \caption{Data from Borexino's low-energy region as presented in \cite{Agostini:2018uly,Borexino_data}. The collaboratioin's best fit to the data is shown overtop the raw data, with the black circles representing the residuals (i.e.\ data-fit). The vertical band shows our binning into low-energy region I and low-energy region II, and the horizontal black lines correspond to the average rate used in each region to set our exclusions. The rate chosen in LER-I lies beneath well understood radioactive backgrounds, but above the data's residuals and the $^7$Be neutrino event rate (not shown). In LER-II the rate is conservatively chosen to lie above the full observed signal. Photon spectral shapes (Dirac $N$ with $\alpha=-1$) are shown for three different masses.    \label{borexino-data}.}
\end{figure}
We consider data from Borexino and Super-Kamiokande to set constraints on a neutrino dipole portal. We perform a conservative  analysis on the Borexino low-energy region (LER) data set from \cite{Agostini:2018uly} (available online at \cite{Borexino_data}), binning their data into a region with $E_\text{vis}<0.8$ keV (LER-I) and a region with $E_\text{vis}>0.8$ keV (LER-II). Since the efficiency of Borexino is so high and photons are reconstructed nearly identically to electrons we neglect efficiencies in our estimates (setting them equal to one). Our signal's photon spectrum is very broad and so we do not include the effects of the $O(50$ keV) broadening from the detector resolution as this will be a negligible effect.

Our analysis is (up to a coarse re-binning) a rate-only analysis and so is insensitive to spectral shape details. In  LER-I the Borexino residuals are small relative to the event rate, with eight different radioactive backgrounds contributing to the expected event rate. We use an event rate of $0.1$ events per 100 $t-$day per keV in LER-I. This corresponds roughly to the full measured event rate in LER-I after subtracting off the $^{210}$Po background and is larger than the $^{7}$Be solar neutrino signal (which Borexino was able to measure with percent level precision). This is important, because the spectrum tends to be rather flat as can be seen in \cref{borexino-data}, which qualitatively resembles the $^{7}$Be signal. A quoted statistical significance would therefore require some prior on the $^{7}$Be flux normalization uncertainty while simultaneously accounting for the tight correlation between the $^{7}$Be signal from $\nu e$ scattering and the $\nu\rightarrow N \rightarrow \nu \gamma$ signal from the solar neutrinos. We do not attempt this here, and instead set limits whenever the rate from $N\rightarrow \nu \gamma$ would be rougly three times as large as the expected Standard Model $\nu e$ scattering signal  with nominal expectations for the solar neutrino flux.  For LER-II we do not have to contend with large radioactive backgrounds, and choose to set constraints by demanding that the $N$ decays do yield more than $0.01$ events per 100 t-day.  As can be seen in \cref{borexino-data} this would completely overwhelm the signal observed in the detector and is clearly incompatible with the measured data. A more sophisticated analysis would include a full spectral shape fit, and an accounting for the daily and seasonal modulation of the signal.
\begin{figure}
    \includegraphics[width=\linewidth]{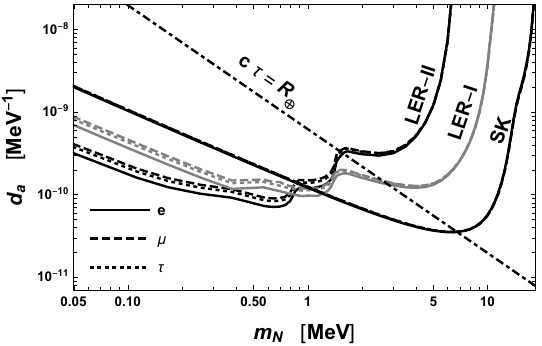}
    \caption{Exclusions from the three experimental data-sets considered in this work for a Majorana-$N$ (box distributed photons). The regions above the lines are excluded. Sensitivity to dipole couplings are largely independent of flavor as can be clearly seen.  The diagonal line shows where the cross-over between $\Ldec \gg R_\oplus$ and $\Ldec\ll R_\oplus$ is expected to take place. We assume a simplified model of the earth with a uniform density, $\overline{n}_A=1.03~\text{cm}^{-3}$, and a $Z_\text{eff}=11.8$. For comparisons to other probes of neutrino-dipole portals see \cref{muon-only}.  \label{Exclusions-tri}}
\end{figure}
For Super-Kamiokande we use recent results from SK-IV, taking Fig.\ 14 of \cite{Abe:2016nxk} \emph{after} cuts were applied. Photon events are difficult to distinguish from electron or positron events since the radiation length of an MeV photon is tens of centimeters, and so pair produces rapidly in medium. The cuts applied on the SK-IV data involved external event vetoes and a tight fiducial volume cut \cite{Abe:2016nxk}. Since the decay length of $N$ is  four to five orders of magnitude larger than the Super-Kamiokande detector, we would expect the photons to be uniformly distributed throughout the detector volume and so these cuts do not modify the expected rate normalized to fiducial volume.  To set limits on $d_\text{eff}$ we require that the $N\rightarrow \nu\gamma$ do not produce a signal in excess of the observed rate above $4$ MeV, which we calculate to be 4.62 events per kton-day. We do not make use of any spectral shape information, day-night asymmetry, or zenith angle dependence all of which could be leveraged for improved sensitivity in a dedicated analysis. 

Finally, we emphasize that the limits set here are  conservative and systematically \emph{underestimate} the reach of both experiments. It is not entirely fair to compare the sensitivities of Borexino and Super-Kamiokande based on the current analysis since each experiment's own detailed spectral shape and day-night asymmetry may significantly impact their ultimate sensitivity; this comparison warrants further study. 
\begin{figure}
    \includegraphics[width=\linewidth]{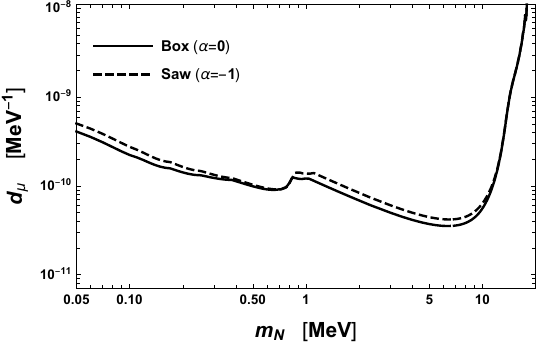}
    \caption{Comparison of the sensitivity for a Majorana ($\alpha=0$) vs Dirac particle with $\alpha=-1$ in the $d-m_N$ plane for an electro- and tau-phobic dipole coupling ($d_e = d_\tau=0$). We see that the sensitivity is relatively insensitive to the shape of the photon spectrum in a rate-only analysis. Constraints are obtained by taking the most stringent of the constraints from Borexino and/or Super-Kamiokande.  \label{Exclusions-box}}
\end{figure}

\section{Conclusions \label{Conclusions}} 

The presence of solar neutrinos and coherent upscattering with the high-density, and moderate-$Z$ material of the Earth's mantle can lead to a dramatic flux of sterile neutrinos passing through terrestrial detectors. Theses fluxes are so large that they compensate for the relatively low probability of decay within the detector volume resulting in rates that dominate over elastic scattering by orders of magnitude. 
\begin{table}
\centering
$\begin{array}{ cccccccc }
\multicolumn{1}{c}{~} &
\multicolumn{2}{c}{\text{[MeV]} } &
\multicolumn{1}{c}{\text{[Events per 100t-day} ]} \\ 
\cmidrule(lr){2-3}
\cmidrule(lr){4-4}
\text{Experiment.} 		& E_\text{min}& E_\text{max}	&  \text{Excluded Rate} \\
\midrule
\text{Borexino LER-I}		& 0.2 	& 0.8  	&  60	\\
\text{Borexino LER-II}		& 0.8   & 2.5	&  17 \\
\text{Super-Kamiokande IV}		& 4.49		& 15.5 & 0.46\\
 \bottomrule
\end{array}$
\caption{ Summary of experimental data used in setting constraints in this paper. Taken from Fig.\ 2 of  \cite{Agostini:2018uly} (available online at \cite{Borexino_data}), and Fig.\ 14 of \cite{Abe:2016nxk}. The excluded rate is a highly conservative choice \emph{c.f.} \cref{borexino-data}. A proper statistical analysis, spectral shape information, and day-night asymmetry could easily improve sensitivity. 
\label{experimental-data}}
\end{table}

As we explore in \cite{Plestid:2020ssy}, similar phenomenology can be used to study other upscattering processes such as $\nu\rightarrow N$ via a mass-mixing portal, followed by $N\rightarrow \nu e^+ e^-$. The phenomenology differs qualitatively from the dipole-scattering example, however, because the upscattering process is quasi-isotropic in the lab frame. This eliminates day-night asymmetry to a first approximation, and requires a slightly re-worked treatment. We refer the interested reader to \cite{Plestid:2020ssy} where these details are discussed and similar bounds are worked out from the non-observation of energetic $e^+e^-$ pairs in Borexino.

Unsurprisingly, luminous neutrino signals share many of the same qualitative features with luminous dark matter. The signal rate is independent of detector composition depending only on the volume of the detector, scattering is preferentially forward, and the signal modulates in time. While similarities exist, there are also important differences.  Luminous dark matter is expected to be heavy $m_\chi \sim 1$ TeV, with a mass-splitting of $O(100~\text{keV})$ and a non-relativisitic velocity \cite{Eby:2019mgs}. This means that the resultant photons are monochromatic with an energy set by the mass-splitting, and that threshold effects are important\footnote{For some parameter space in \cite{Eby:2019mgs} upscattering into the excited state is only possible if the target nucleus is very heavy e.g.\ $^{208}$Pb.}.  In contrast the model considered in this paper leads to moderately relativistic sterile neutrinos, whose lab-frame photon distribution is therefore broadened. The solar neutrino flux is broad, and since neutrinos are much lighter than nuclei, the only threshold effect that manifests itself in this work is the requirement that $E_\nu \geq m_N$.  Looking forward it is important to emphasize that \emph{any} astrophysical source of neutrinos (e.g.\ atmospherics) could produce similar phenomenology, albeit with the potential mandatory inclusion of neutrino oscillation effects. Luminous signals of inelastic transitions within the Earth have many common ingredients, but their signatures inside detectors can be somewhat model dependent. 
Our study further motivates a dedicated program at future large scale detectors to search for upscattered new physics within the Earth. This includes inelastic dark matter, but also any neutrinophilic inelastic transition.

Our study has demonstrated the discovery potential using solar neutrinos, but heavier sterile neutrinos, $N$, can be produced from atmospheric neutrinos and it would be interesting to study their impact on dipole portals and other inelastic upscattering operators.  Atmospheric neutrinos have a much smaller flux, but could probe heavier $m_N$, where constraints are lacking as shown in \cref{muon-only}. 

Future experiments such as JUNO and Hyper-Kamiokande can serve as sensitive probes of upscattered particles, as can any other large-volume detector capable of detecting MeV scale photons (this statement is independent of detector density).  Searches could leverage time modulation, which needs to be studied in more detail as has been done for luminous dark matter \cite{Eby:2019mgs},  zenith-dependence, and spectral shape characteristics to distinguish new physics from backgrounds.

\section{Acknowledgements}
  I would like to thank Joachim Kopp, Vedran Brdar,  Gordan Krnjaic,  and Matheus Hostert for helpful discussions, and I am especially grateful to Pedro Machado,  Volodymyr Takhistov, Kevin Kelly, and Ciaran Hughes for detailed comments and feedback on early versions of this manuscript. I would like to thank the Fermilab theory group for their hospitality and welcoming research atmosphere.  This work was completed while visiting Fermilab with support from the Intensity Frontier Fellowship. This work was supported by the U.S. Department of Energy, Office of Science, Office of High Energy Physics, under Award Number DE-SC0019095. This manuscript has been authored by Fermi Research Alliance, LLC under Contract No. DE-AC02-07CH11359 with the U.S. Department of Energy, Office of Science, Office of High Energy Physics.

 \vfill 
 \pagebreak
\appendix

\section{Calculation of $\mathcal{I}(\zeta)$ \label{I-calc}}

To calculate the decay-weighted average path length $\mathcal{I}(\zeta)$ we take our detector to be described by a point 
\begin{equation}
    p(t)=(\sqrt{R_\oplus^2-z_0^2}\cos\Omega t, \sqrt{R_\oplus^2-z_0^2}\sin\Omega t,z_0)~,
\end{equation}
defined in the reference frame of the Earth ($\hat{z}$ parallel to the magnetic pole).  We then rotate the Earth's pole via a the rotation matrix 
\begin{equation}
    R_\text{tilt}= \left(
    \begin{array}{ccc}
     \cos \theta_\text{tilt}  & 0 & \sin \theta_\text{tilt} \\
     0 & 1 & 0 \\
     -\sin  \theta_\text{tilt} & 0 & \cos \theta_\text{tilt}  \\
\end{array}
\right)~,
\end{equation}
where $\theta_\text{tilt}=23.5$ degrees. This gives us an Earth that is titled and spinning about its own axis, but is static relative to the Sun. In reality the Earth precesses relative to the Sun. We therefore apply a final rotation $R_\text{prec.}(\omega t)$ for rotation about the $\hat{z}'$ axis (perpendicular to the Sun) ultimately arriving at the depiction shown in \cref{Earth-sketch}
\begin{equation}
    p'(t)=R_\text{prec}(\omega t) R_\text{tilt} p(\Omega t)~.
\end{equation}
Next, we draw a straight line (ray) from $p'$ to the Sun,  which in the solar-beam frame is equivalent to adding $-L \hat{x}$  to $p'$.  We then solve for the value of $L$ for which 
\begin{equation}
    \abs{p'(t)-L \hat{x}}^2= R_\oplus^2~,
\end{equation}
which is the distance to the point of intersection of the ray with the surface of the Earth closest to the Sun. The value of $L$ is then the time-dependent slab length 
\begin{equation}
    L_\text{slab}(\omega t, \Omega t)= L(t) \qq{if} L(t)\geq 0. 
\end{equation}
For a given day of the year the day-averaged rate one should make the replacement $\Omega t \rightarrow \omega t + \Omega t$ (accounting for the slow-drift in position from the Earth's precession relative to the Sun) is given by 
\begin{equation}
     \mathcal{I}_i(\zeta)= \frac{1}{2\pi}\int_0^{2\pi} \dd ~[\Omega t] \qty[1-\e^{-L_\text{slab}/\Ldec}]~,
\end{equation}
with $\omega t= 2\pi \times (\text{day}_i/365)$. For the year-averaged quantity used for limit setting in this paper we have 
\begin{equation}
    \mathcal{I}(\zeta)= \frac{1}{4\pi^2}\int_0^{2\pi}\int_0^{2\pi} \dd~ [\Omega  t] \dd~[ \omega t] \qty[1-\e^{-L_\text{slab}/\Ldec}]~.
\end{equation}

Finally, in a crude two-region model of the Earth's interior, where there is a sharp boundary between the core and mantle at some radius $r < R_\oplus$, the same basic machinery can be used to estimate the time-averaged value of $\overline{n}_A$ and $Z_\text{eff}$ (which, as discussed above, will generally be time dependent for a heterogeneous model of the Earth). Let us take $p'(t)-L\hat{x}$ again, but now search for solutions of the form 
\begin{equation}
    \abs{p'(t)-L \hat{x}}^2= r^2~,
\end{equation}
for which there will generically be two solutions $L_\pm(t)$ corresponding to when the ray goes enters and exits the core. Taking the difference of these two quantities then yields the path-length that resides within the core, which we will denote $L_\text{core}$. The mantle length can then be found from the full slab-length as $L_\text{mantle} = L_\text{slab} - L_\text{core}$. As we will discuss in the next section, for the $\Ldec \gg R_\oplus$ limit this information is sufficient to estimate the added flux due to the high-density, high-$Z$ material of the Earth's core. 

\begin{table}
\centering
$\begin{array}{ cccc}
\text{Molecule.} 		& n ~[10^{22}~ \text{cm}^{-3}]  & \Sigma Z^2 & n \times \Sigma Z^2 ~ [10^{24} ~\text{cm}^{-3}]  	 \\
\midrule
\text{SiO}_2 & 1.81 &  324 & 5.85 \\ 
\text{MgO }& 2.33 &  208 & 4.85 \\ 
\text{FeO} & 0.28 &  740 & 2.04\\ 
\text{CaO} & 0.14 &  464  & 0.64 \\ 
\text{Cr}_2\text{O}_3  & 0.01 & 1344 &  0.12 \\
\text{Ni}\text{O} & 0.01 &  848 & 0.07 \\ 
 \bottomrule
\end{array}$
\caption{Mantle composition calculated using Table 3.\ of  \cite{WORKMAN200553} and a density of $\overline{\rho}=4$ g/cm$^3$. Elements with $n_i \leq 7 \cdot 10^{19}~\text{cm}^{-3}$ are omitted. 
\label{earth-table}}
\end{table}

\section{Effective composition of the Earth \label{Earth-Material}}

Our choice of parameters for the uniform-density model of the Earth was driven by conservative considerations. First, the core of the Earth will always yield a higher-rate of $\nu\rightarrow N$ production than an equivalent volume filled with mantle, and so treating the core as if it were made of mantle will \emph{always} underestimate the rate of $N$ production.  The average density of Earth is roughly 5.5 g/cm$^3$; we conservatively use $\overline{\rho}=4$ g/cm$^3$ in setting our limits since for certain regions of parameter space we consider only that outer mantle contributes significantly to upscattering. 

For shorter decay lengths probed by SK (namely $\lambda \lesssim 1$ km) one may worry that  $\overline{\rho}=4$ g/cm$^3$ is not a conservative choice since the crust has a slightly lower density of roughly $2.7$ g/cm$^3$. This concern turns out to be unwarranted, however, because in this limit both the crust beneath and above the experiment (i.e\ the overburden) would contribute to scattering roughly doubling the signal over the ``night only'' estimates presented here. We therefore conclude that $\overline{\rho}=4$ g/cm$^3$ is indeed conservative over the full range of parameter space we consider. 

The elemental composition of the Earth is given in Table 3.\ of \cite{WORKMAN200553} and we use the DMM column, which gives the mass percentage by molecular compound.  We are ultimately interested in 
\begin{equation}
    \overline{n}_A Z_\text{eff}^2 = \sum_{i} n_i \qty[\sum Z^2]_i ~,
\end{equation}
where the bracketed sum is adding up $Z^2$ within the atoms of each molecular compound.  This can be re-expressed in terms of the mantle density via
\begin{equation}
    n_i = \frac{\overline{\rho}}{m_i} f_m ~,
\end{equation}
where $f_m$ is the mass fraction given in Table 3. of \cite{WORKMAN200553} (Bulk DMM). We can define $Z_\text{eff}^2$ via 
\begin{equation} 
    Z_\text{eff} := \frac{\langle Z^2 \rangle }{\langle Z \rangle} ~, 
\end{equation}
which is a density independent definition. The effective number density is then defined as 
\begin{equation}
   \overline{n}_A= \frac{ \sum_{i} n_i \qty[\sum Z^2]_i }{Z_\text{eff}^2} ~.
\end{equation}

As we have shown above, for $m_N\leq 5-10$ MeV, the combined sensitivity of Borexino and Super-Kamiokande probes parameter space where $\Ldec \gg R_\oplus$. For internal consistency, we do not make use of the high-density, high-$Z$ nature of the Earth's core in the main text, because for the higher mass regions $m_N \geq 1-10$ MeV (depending on the experiment) the parameter space being probed satisfies $\Ldec \ll R_\oplus$. In this limit, any upscattered particles from the Earth's core would decay prior to reaching the detector and so, for limit setting purposes, including the Earth's core in the definition of the average column density is unrealistically aggressive. 

\begin{figure}
     \includegraphics[width=\linewidth]{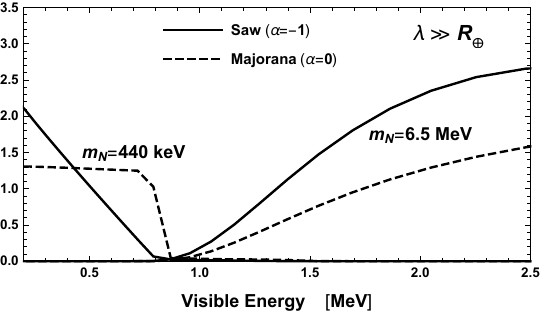}
     \caption{Dirac with $\alpha=-1$ vs Majorana spectral shape in the energy corresponding to the SK-IV solar neutrino dataset for a relevant benchmark mass. The condition $\Ldec\ll R_\oplus$ is assumed corresponding to a value of $d_\text{eff}$ above the dashed line in  \cref{Exclusions-tri,Exclusions-box}.  } 
\end{figure}

For the low-mass region $m_N \lesssim 2$ MeV, however,  it is easy to include the effect of the Earth's core, since in this region $\Ldec \gg R_\oplus$, most all of the upscattered particles will survive to the detector,  and the core just serves to enhance the effective column density.  The tricky part to include is the geometry of the core, which only appears along the line-of-sight to the Sun in the winter months the details  of which are detector latitude dependent (see e.g.\ Fig.\ 3 of \cite{Eby:2019mgs}). Using the procedure outlined at the end of \cref{I-calc} we estimate than for a two-component core-mantel model with $r=0.5 R_\oplus$ that the $\langle L_\text{core}/L_\text{slab}\approx 0.2 \rangle$ i.e.\ that in a full year, of the distance traversed by solar neutrinos en route to the detector, roughly twenty percent lies inside the Earth's core. The added flux from the core of the Earth can then be found by taking a weighted average of $n_A Z_\text{eff}^2$ in the core and mantle. Using a density for the core of $13$ g/cm$^3$ and 4 g/cm$^3$ for the mantle (yielding an average density of $5.125$ g/cm$^3$ for the Earth), and treating the core as $90\%$ iron $10\%$ oxygen by mass ($Z_\text{eff}=24.1$ and $\overline{n_A}=1.54 \cdot 10^{23} ~\text{cm}^{-3}$). We find that the signal is enhanced by a factor of 
\begin{equation}
  \frac{  0.8 [ \overline{n}_A Z_\text{eff}^2]_\text{mantle} + 0.2 [ \overline{n}_A Z_\text{eff}^2]_\text{core}}{\overline{n}_A Z_\text{eff}^2}\approx 2.1 ~.
\end{equation}
The effect of the core would serve to roughly double the year-averaged rate. 

\begin{figure}
    \includegraphics[width=\linewidth]{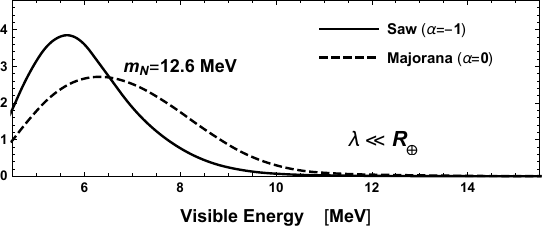}
    \caption{Dirac (with $\alpha=-1$) vs Majorana spectral shape in the energy corresponding to the Borexino LER for two benchmark masses that can be probed using that dataset. The condition $\Ldec\gg R_\oplus$ is assumed so as not to be excluded \emph{c.f.} \cref{Exclusions-tri,Exclusions-box}.} 
\end{figure}

\section{Dirac vs.\ Majorana}

In \cite{Berryman:2019dme,Kayser:2018mot,Balantekin:2018azf,Balantekin:2018ukw} the authors discuss the implications of a new physics discovery whose progenitor is a decaying sterile neutrino. As discussed in this paper for $N\rightarrow \nu \gamma$, the  kinematic distribution of daughter particles is different depending on whether $N$ is Dirac or Majorana and relative size and phases of the magnetic and electric dipole moments \cite{Kayser:2018mot,Balantekin:2018azf,Balantekin:2018ukw}. Thus,  if an anomalous photon yield were to be discovered in a large volume detector, the energy spectrum of photons could be used to infer whether or not $N$ is Dirac as opposed to Majorana.

In this appendix we provide a plot comparing the Dirac (with $\alpha=-1$) vs Majorana (box vs tri) spectral shapes. These shapes arise from a combination of the solar neutrino spectrum and the rest-frame decay properties of $N$.  One also must understand the hierarchy of $\Ldec$ and $R_\oplus$, which introduces additional energy dependence in the spectrum.

\vfill

\pagebreak
\section{Solar Neutrino Flux}

\begin{figure}[hb]
    \includegraphics[width=0.9\linewidth]{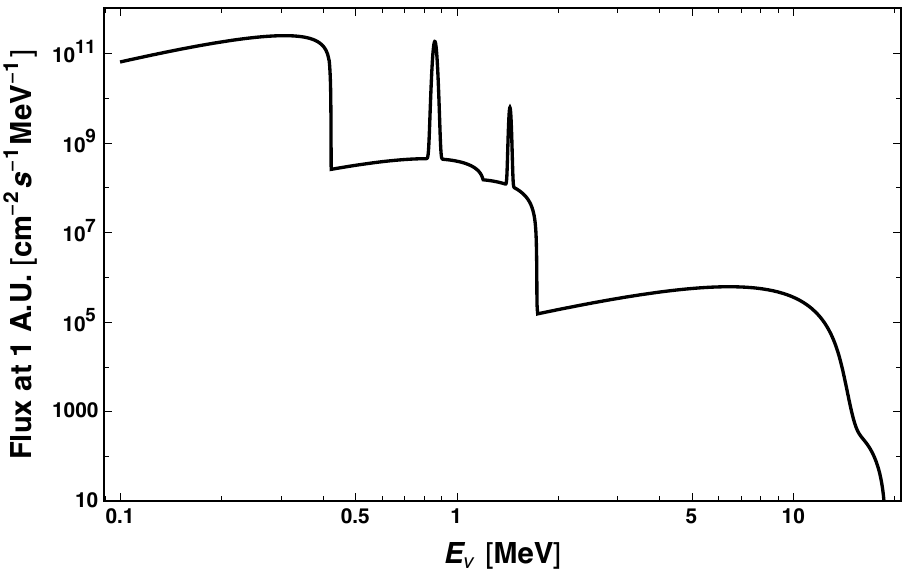}
    \caption{Solar neutrino flux.  Shapes are taken from \cite{Bahchall_url} and normalizations from Tab.\ 2 of \cite{Serenelli:2011py} (AGSS09 \cite{Asplund:2009fu}).\label{Solar-Flux} }
\end{figure}

For completeness we describe our input solar neutrino flux. This is most easily summarized in visual form in \cref{Solar-Flux}. The $^7$Be line at 384 keV has been included but is sub-dominant to the pp and hence not visible. We treat the. $pep$ and $^7$Be lines as Gaussians with a width of 10 keV. 

\bibliography{biblio.bib}
\end{document}

%% file: Commands/ryansCommands.tex
\def\e{\mathrm{e}}
\def\Ldet{\ell}
\def\Ldec{\lambda}